\documentclass[aps,prb,twocolumn,superscriptaddress,showpacs]{revtex4-1}

\usepackage{graphicx}
\usepackage{calc}
\usepackage{bm}
\usepackage{color}
\usepackage{textcomp}

\begin{document}

\title{Josephson diode and spin-valve effects on the surface of altermagnet CrSb}

\author{V.D.~Esin}
\author{D.Yu.~Kazmin}
\author{Yu.S.~Barash}
\author{A.V.~Timonina}
\author{N.N.~Kolesnikov}
\affiliation{Institute of Solid State Physics of the Russian Academy of Sciences, Chernogolovka, Moscow District, 2 Academician Ossipyan str., 142432 Russia}
\author{E.V.~Deviatov}
\affiliation{Institute of Solid State Physics of the Russian Academy of Sciences, Chernogolovka, Moscow District, 2 Academician Ossipyan str., 142432 Russia}
\affiliation{V.L. Ginzburg Research Centre for High-Temperature Superconductivity and Quantum Materials, P.N. Lebedev Physical Institute of RAS, Moscow 119991, Russia}

\date{\today}

\begin{abstract}
We experimentally investigate charge transport in In-CrSb and In-CrSb-In proximity devices, which are formed as junctions between superconducting indium leads and thick single crystal flakes of altermagnet CrSb. For double In-CrSb-In junctions, $dV/dI(B)$ curves are mirrored in respect to zero field for two magnetic field sweep directions, which is characteristic behavior of a Josephson spin valve. Also, we demonstrate Josephson diode effect by direct measurement of the critical current for two opposite directions in external magnetic field. We interpret these observations as a joint effect of the spin-polarized topological surface states and the altermagnetic spin splitting of the bulk bands in CrSb. For a single In-CrSb interface, the superconducting gap oscillates in magnetic field for both field orientations, which strongly resembles the Fulde-Ferrell-Larkin-Ovchinnikov (FFLO) behavior. FFLO is based on finite-momentum Cooper pairing, therefore, it is fully compatible with the requirements for the Josephson diode effect. 
\end{abstract}


\maketitle

\section{Introduction}

In altermagnets, the concept of spin-momentum locking~\cite{Armitage,sm-valley-locking} was extended to the case of weak spin-orbit coupling, i.e. to the non-relativistic groups of magnetic symmetry~\cite{alter_common,alter_mazin}. As a result, the small net magnetization is accompanied by 
alternating spin splitting in the k-space~\cite{alter_common,alter_Kramers,alter_josephson}. In the simplest case of d-wave symmetry, the  up-polarized subband can be obtained by $\pi/2$ rotation of the down-polarized one in the k-space~\cite{alter_supercond_notes,alter_normal_junction}. Consequently, an altermagnet sometimes behaves as an antiferromagnet, and sometimes as a ferromagnet, depending on the crystal-field or interface-crystal relative orientations.

There are many theoretical predictions on possible effects of  proximity-induced superconductivity in  
altermagnets~\cite{alter_supercond_notes,alter_supercond_review1,proxi1,proxi2,proxi3,proxi4,alter_josephson,alter_josephson1,SN1,SN2,SN3,SN4,alterSN}. For example, as the direct consequence of k-dependent spin splitting, orientation-dependent effects can be expected for different 
superconductor-altermagnet-superconductor~\cite{alter_josephson,alter_josephson1} (SNS) and 
superconductor-altermagnet~\cite{alter_supercond_review1,esin_physicaB2025,SN1,SN2,SN4,alterSN} (SN) structures. For applications, the absence 
of stray fields in altermagnets makes them advantageous for superconducting spintronics logic circuits.

Spin-momentum locking is a key feature not only of altermagnets, but also of a large class of topological materials~\cite{Armitage}. In the latter case, it is responsible for the spin polarization of the  topological surface states~\cite{PhysRevB.100.195134}, which are able to carry supercurrents over extremely large  distances~\cite{topojj1,topojj2,topojj3,topojj4,topojj5}. Weyl semimetal states have also been theoretically predicted in altermagnets~\cite{AMtopology1,AMtopology2}. In proximity topological devices, spin-polarized surface states lead to the different realizations of the Josephson diode effect (JDE). 

The diode effect in superconductors occurs if the critical current $I_c$ is different for two opposite directions. For the Josephson diode effect, the absolute values of $I_c^+(B)$ (positive $I_c$) and $I_c^-(B)$ (negative $I_c$) differ for the two current  directions (nominally + and - ones). This behavior should be distinguished from usual difference between the critical and the return currents owing to the finite capacitance of devices, while JDE emerges under certain conditions in systems with broken time-reversal and inversion symmetries~\cite{infgt,JDE,JDEGen,JDE2,JDERev1,JDEComm,alter_supercond_review1,JDERev2,JDERev3,JDERev4,aunite}. 

As possible physical mechanisms of JDE, Cooper pairs can acquire a finite momentum and give rise to a diode effect in superconductors with strong spin-orbit coupling~\cite{JDE16,JDE17,JDE18}. In paramagnetic and centrosymmetric Dirac semimetal NiTe$_2$, the finite momentum pairing results from the momentum shift of topological surface states under an in-plane magnetic field due to the spin-momentum locking~\cite{JDE,aunite}. In superconducting heterostructures with non-coplanar magnetization textures, breaking the magnetization reversal symmetry can result in the direct coupling between the magnetic moment and the supercurrent, and, therefore,  in the Josephson diode effect~\cite{buzdin2008,buzdin2009,linder2014,bergeret2017,buzdin2017,ajam2024}. 

The Josephson diode effect is typically observable for Josephson spin-valves (JSV), where the ferromagnetic multilayer~\cite{valve1,valve2} is sandwiched between two superconducting electrodes in vertical~\cite{reverse,jsv1,jsv2,jsv3,jsv4,krasnov,jsv6,jsv7} or planar geometries~\cite{golubov}. While in conventional Josephson junctions supercurrent is modulated by magnetic flux, in a Josephson spin valve it is mainly controlled by the relative orientation of magnetic layers, giving rise to the critical current asymmetry and reversal. The magnetic topological materials demonstrate spin-valve transport properties~\cite{timnal,cosi,bite,gete}, since the spin-polarized surface states and the  ferromagnetic bulk constitute two spin-polarized systems. Thus, JSV can also be realized on the surface of the proximized magnetic topological semimetal~\cite{infgt}.  Also, the Josephson spin-valve behavior has recently been observed in Nb-Mn$_3$Ge-Nb junctions containing a single interlayer of the topological chiral antiferromagnet Mn$_3$Ge~\cite{jsv5}. 

\begin{figure}
\includegraphics[width=\columnwidth]{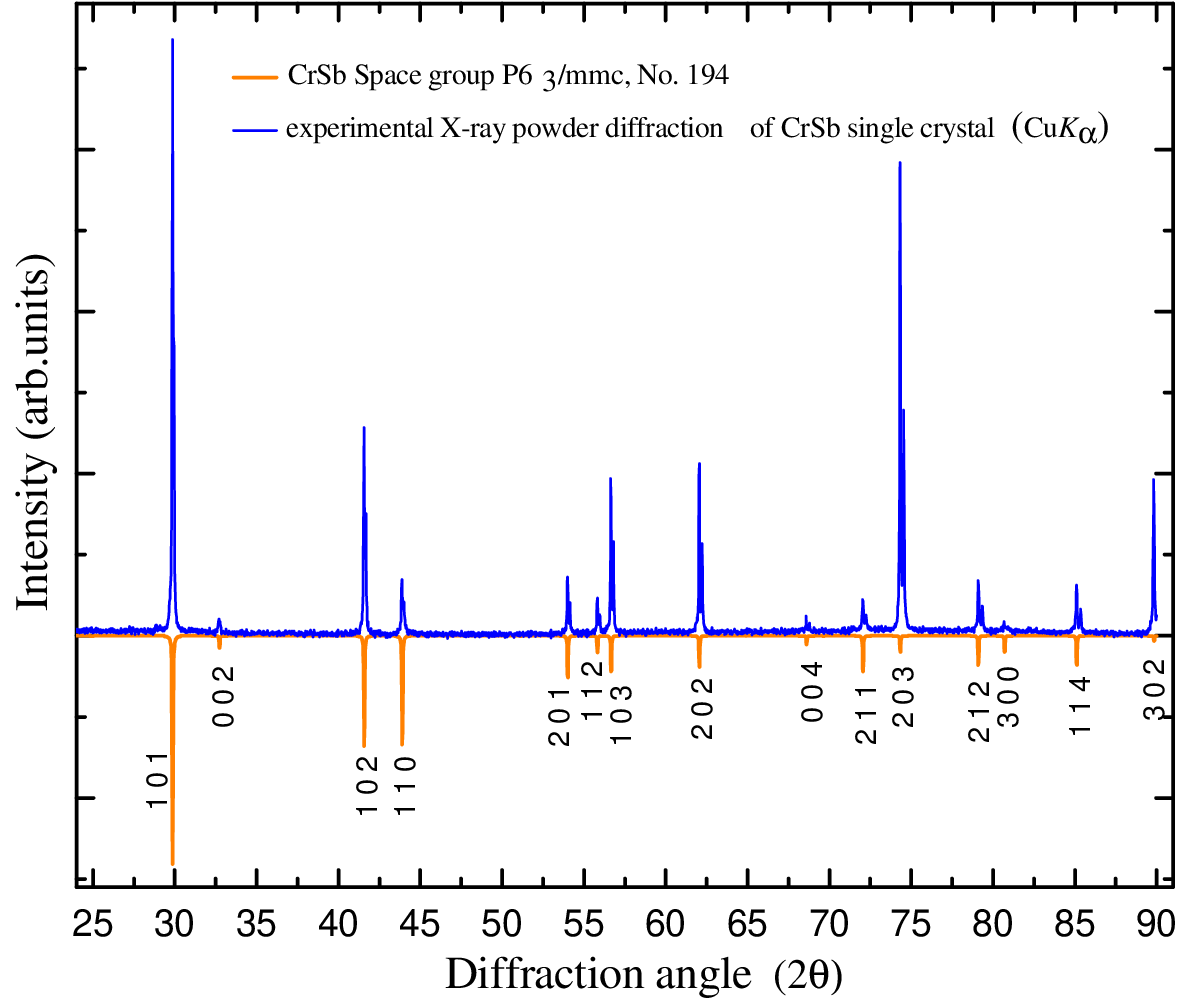}
\caption{(Color online)  The X-ray powder diffraction  pattern (Cu K$_{\alpha}$ radiation), which is obtained for the crushed CrSb single crystal. The single-phase  CrSb is confirmed with the space group $P6_3 /mmc$ No. 194. }
\label{fig0}
\end{figure}

The Josephson diode effect has recently  been predicted for junctions incorporating altermagnets~\cite{AMJDE1,alter_supercond_review1,AMJDE2,AMJDE3}. Experimental investigations of the Josephson current asymmetry  can be conveniently performed for CrSb, which has recently been identified as a new altermagnetic metal through spin-integrated soft  X-ray angular-resolved photoelectron spectroscopy (SX-ARPES)~\cite{ARPES1_CrSb}. In contrast to the  well-known altermagnet MnTe~\cite{satoru,Dichroism,MnTe_SO}, spin-orbit coupling plays a minor role in the low energy band structure in CrSb, so  the altermagnetism  is  well defined and characterized by non-relativistic spin-group symmetries~\cite{ARPES2_CrSb}. Also,  CrSb metal is of high bulk conductance even at low temperatures, which facilitates fabrication of transparent SN interfaces for the altermagnetic-based proximity devices in comparison with the  semiconductor  MnTe~\cite{esin_physicaB2025}.

 CrSb reveals both altermagnetic and topological features~\cite{Weyl alter2_CrSb}.  Topological surface states have been clearly demonstrated on the  (100) cleaved side surface close to the Fermi level originating from bulk band topology in CrSb~\cite{Weyl alter1_CrSb}. It was also confirmed by observation of the interplay between the altermagnetic  bulk and the topological surface magnetizations~\cite{crsbsbs}. Thus, it is reasonable to investigate possible anomalies in Josephson effect induced by these surface states for the altermagnetic candidate CrSb. 

Here, we experimentally investigate charge transport in In-CrSb and In-CrSb-In proximity devices, which are formed as junctions between superconducting indium leads and thick single crystal flakes of altermagnet CrSb. For double In-CrSb-In junctions, we demonstrate the characteristic behavior of a Josephson spin valve and Josephson diode effects by direct measurement of the critical current in external magnetic field.  For a single In-CrSb interface, the superconducting gap oscillates in magnetic field for both field orientations,  before its full suppression.

\section{Samples and technique}

\begin{figure}
\includegraphics[width=\columnwidth]{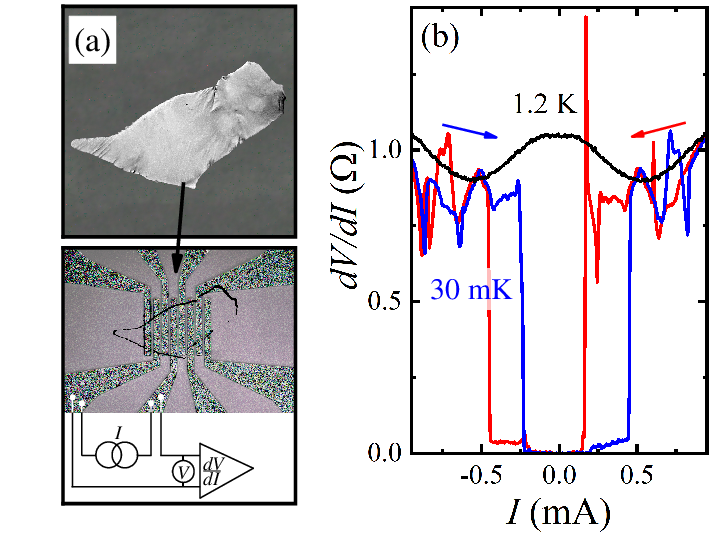}
\caption{(Color online) (a) Optical image of the CrSb single-crystal flake (top image), which is placed on 
the pre-defined In leads pattern (the bottom one) to form 2~$\mu$m separated In-CrSb junctions.  To investigate Josephson current through the CrSb surface,  In-CrSb-In resistance is measured in a standard four-point technique. (b) Josephson $dV/dI(I)$ curves at 30~mK temperature, the blue and the red curves  are for two opposite current sweep directions, respectively. Zero-resistance state can be clearly seen, which is accompanied by usual hysteresis for the critical and the return currents. The black curve demonstrates suppression of the Josephson effect at 1.2~K. The curves are obtained after sample cooling in zero magnetic field, before any sample magnetization.}
\label{fig1}
\end{figure}

CrSb single crystals were synthesized by reaction of pure Cr (99.996\%) and Sb (99.9999\%), which were mixed in the stoichiometric ratio and then heated in an evacuated silica ampule up to 1000$^\circ$C with the rate of 15$^\circ$C/h in a gradient-free furnace. The load was held at 1000$^\circ$C for 72 hours and then cooled down slowly (11$^\circ$C/h) to the room temperature. The crystals grown are faceted single crystals with the space group $P6_3 /mmc$ No. 194 and the stoichiometric composition, as confirmed by X-ray diffraction analysis, see Fig.~\ref{fig0}.

The images in Fig.~\ref{fig1}(a) shows schematically sample fabrication.   The desired experimental geometry is defined by indium leads pattern on a standard Si/SiO$_2$ substrate. The 5~$\mu$m wide In leads are separated by 2~$\mu$m intervals, as depicted in Fig.~\ref{fig1} (a). They are formed by lift-off technique after thermal evaporation of 100~nm thick In. 

CrSb single crystal flakes are obtained  by mechanical exfoliation. CrSb is a three-dimensional altermagnet, therefore, one has to select  relatively thick (above 1~$\mu$m) single crystal flakes with $\approx$100~$\mu$m lateral size. The selected flake is placed on the In contact leads.  After initial single-shot pressing by another oxidized silicon substrate,  the flake is firmly connected to the In leads. This procedure provides Andreev In-CrSb junctions (about 1~Ohm normal resistance), stable in subsequent cooling cycles, which has been verified for a wide range of materials~\cite{inwte1,topojj4,infgt,ingete}. As an additional advantage, the In-CrSb interfaces are protected from any contamination by CrSb bulk and the Si/SiO$_2$ substrate. 

Every In-CrSb junction can be independently characterized in a three-point connection scheme: one In contact is grounded, the neighbor (2~$\mu$m separated) In contact is used as a voltage probe, while current is fed through another (remote) contact.  We use an additional (fourth) wire to the grounded contact.  Thus all the wire resistances are excluded, which is necessary for low-impedance samples.  If there is no supercurrent 
between two neighbor In leads,  the potential probe mostly reflects the voltage drop across the grounded junction~\cite{topojj2,aunite}, i.e. the Andreev reflection~\cite{andreev,tinkham} at the In-CrSb  interface. Otherwise, if the voltage drop is exactly zero for some current range, Josephson current connects the two neighbor In leads. In the latter case, the In-CrSb-In resistance is measured in a standard four-point technique~\cite{inwte1,topojj4,infgt,ingete,aunite}, as schematically presented in Fig.~\ref{fig1} (a). 

To obtain differential $dV/dI(V)$ and $dV/dI(I)$ characteristics, dc current is additionally modulated by a low (100~nA) ac component, thus, the lock-in detected ac voltage is proportional to the differential resistance $\sim dV/dI$. The signal is confirmed to be independent of the modulation frequency within 100 Hz -- 10kHz range, which is defined by the applied filters.  The measurements below are performed for 30~mK and 1.2~K temperatures in a dilution refrigerator.

\section{Experimental results}

\subsection{Double In-CrSb-In junctions} \label{double}

\begin{figure}
\includegraphics[width=\columnwidth]{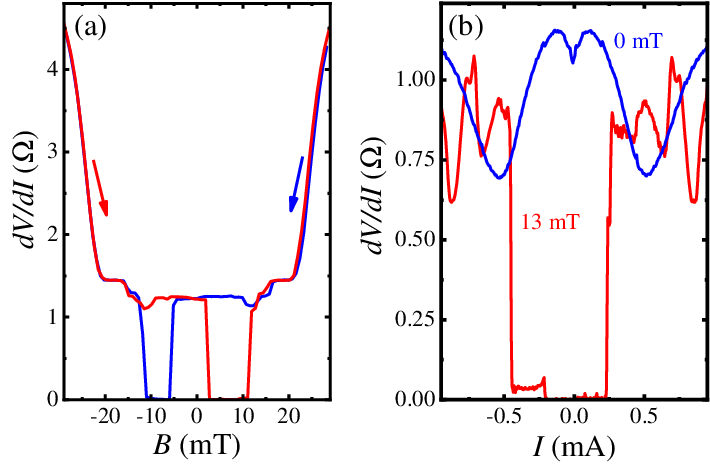}
\caption{(Color online) (a) Josephson spin-valve effect as the $dV/dI(B)$ curves reversal for two opposite magnetic field sweep directions at 30~mk temperature. The zero-resistance  region is not only shifted to finite magnetic fields, but the $dV/dI(B)$ curves are mirrored in respect to zero field, including regions with finite resistance. 
(b) $dV/dI(I)$ curves after magnetization of the sample. Differential resistance is always finite at zero field now, but there is wide  
zero-resistance region at 13~mT magnetic field. Magnetic field is directed normally to the In-CrSb interfaces (i.e. to the CrSb flake).
}
\label{fig2}
\end{figure}

For two transparent neighbor In-CrSb junctions, Josephson current connects the indium leads, see 
Fig.~\ref{fig1} (b). The zero-resistance  region is slightly asymmetric for every current sweep direction,  
so there is usual  hysteresis for the critical and the return currents. The current range is much below the critical 
current of the indium leads, which we can estimate for our leads' dimensions as about 
30 mA for the known~\cite{indium-current} value $j\approx 3\times 10^6$ A/cm$^2$.  

Josephson current is suppressed at 1.2~K temperature in Fig.~\ref{fig1} (b), so the $dV/dI(I)$ curve is of 
standard Andreev shape~\cite{tinkham} with two resistance minima at $\approx\pm 0.5$~mA current bias. While 
the normal resistance value is $\approx 1\Omega$ in Fig.~\ref{fig1} (b), the superconducting gap can be 
estimated as 0.5~mA$\times$~1~$\Omega \approx 0.5$~meV, which well corresponds to the  bulk indium 
gap~\cite{indium-gap-field}.  

The curves in Fig.~\ref{fig1} (b) are obtained after sample cooling in zero magnetic field. Fig.~\ref{fig2} (a) shows the effect of the magnetic field for two opposite field sweep directions. While sweeping from $\pm$100~mT magnetic field, the zero-resistance regions are shifted to finite fields $\approx\mp$10-13 mT, as confirmed by  the $dV/dI(I)$ curves in Fig.~\ref{fig2} (b): the differential resistance is finite at zero field value, but there is zero-resistance region at 13~mT magnetic field. The field is directed normally to the In-CrSb interfaces (i.e. to the CrSb plane), the values are well below the indium critical field (110 mT for the 100 nm thick films~\cite{indium-gap-field},  which we confirm for our indium leads). Also, there is no noticeable  frozen flux for this field range for our solenoid. 

The main experimental finding is the $dV/dI(B)$ curves reversal in Fig.~\ref{fig2} (a). Indeed, for two magnetic field sweep directions,  
the $dV/dI(B)$ curves are mirrored in respect to zero field, including regions with finite $dV/dI$ differential resistance. The observed 
behavior is known for Josephson spin valves~\cite{infgt,jsv5}. 

\begin{figure}
\includegraphics[width=\columnwidth]{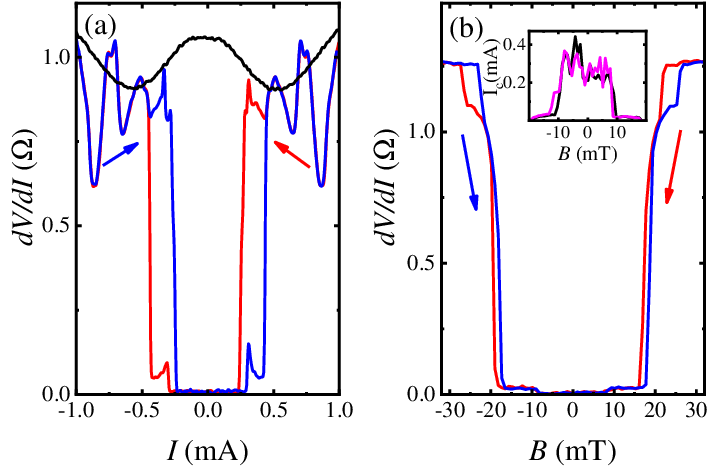}
\caption{(Color online) Josephson spin-valve effect for the parallel to the In-CrSb interfaces magnetic field orientation at 30~mK temperature. 
(a)  Before magnetization of the sample, $dV/dI(I)$ curves well reproduce ones from  Fig.~\ref{fig1} (b). The zero-resistance  region is slightly asymmetric for every current sweep direction (red and blue curves), it is suppressed at 1.2~K temperature (the black one).
(b) After magnetization of the sample, the $dV/dI(B)$ curves are  mirrored in respect to zero field 
(see also the regions of finite resistance), confirming the Josephson spin valve behavior. 
Inset shows the Josephson diode effect as another demonstration of the $dV/dI(B)$ curves reversal: the measured critical currents $I_c^\pm(B)$  are asymmetric in respect to zero field, but they coincide well when drawn as $I_c^+(B)$ (positive $I_c$) and $-I_c^-(-B)$ (the inverted negative $I_c$ for the inverted field value)~\cite{infgt,JDE,aunite}.}
\label{fig3}
\end{figure}

The spin-valve effect is not so strong in parallel to the In-CrSb interfaces magnetic field orientation, see Fig.~\ref{fig3}. After sample cooling in zero field, $dV/dI(I)$ curves in Fig.~\ref{fig3} (a) well reproduce those from Fig.~\ref{fig1} (b): the zero-resistance  region is slightly asymmetric for every current sweep direction, it is suppressed at 1.2~K temperature in Fig.~\ref{fig3} (a). While sweeping the magnetic field, 
 the $dV/dI(B)$ curves are clearly mirrored in respect to zero field (see also the regions of finite resistance), confirming the Josephson spin-valve behavior. Thus, the spin-valve effect is sensitive to the $\pi/2$ rotation of the magnetic field, which resembles the results for CrSb  magnetization~\cite{crsbsbs}.

Since the Josephson spin valves usually demonstrate the Josephson diode effect~\cite{infgt,JDE,aunite}, we show the $I_c^+(B)$ and the inverted  $-I_c^-(-B)$ critical current values in the inset to Fig.~\ref{fig3} (b). The Josephson critical current $I_c^\pm$ is measured for the transition from the superconducting $dV/dI= 0$ state to the resistive $dV/dI > 0$ state for the positive and negative dc currents  (i.e. for  + and - current sweep directions).  Since the transition from the superconducting to resistive states is stochastic, to obtain $I_c$ with high accuracy, we sweep the dc current ten times from zero value to some value well above the critical current $I_c$  at fixed $B$ and then determine $I_c$ for this field $B$ as an average value of $dV/dI$ breakdown positions. The obtained $I_c^\pm(B)$ curves are asymmetric in respect to zero magnetic field in the inset to Fig.~\ref{fig3} (b), but they coincide only when drawn as $I_c^+(B)$ (positive $I_c$) and $-I_c^-(-B)$ (the inverted negative $I_c$ for the inverted field value), which is the direct demonstration of the Josephson diode effect~\cite{infgt,JDE,aunite}.

As a result, double In-CrSb-In not only transfer Josephson current at millikelvin temperatures on the surface of altermagnet CrSb, but also demonstrate prominent Josephson spin-valve and diode effects in external magnetic field.

\subsection{Single In-CrSb junction}

\begin{figure}
\includegraphics[width=\columnwidth]{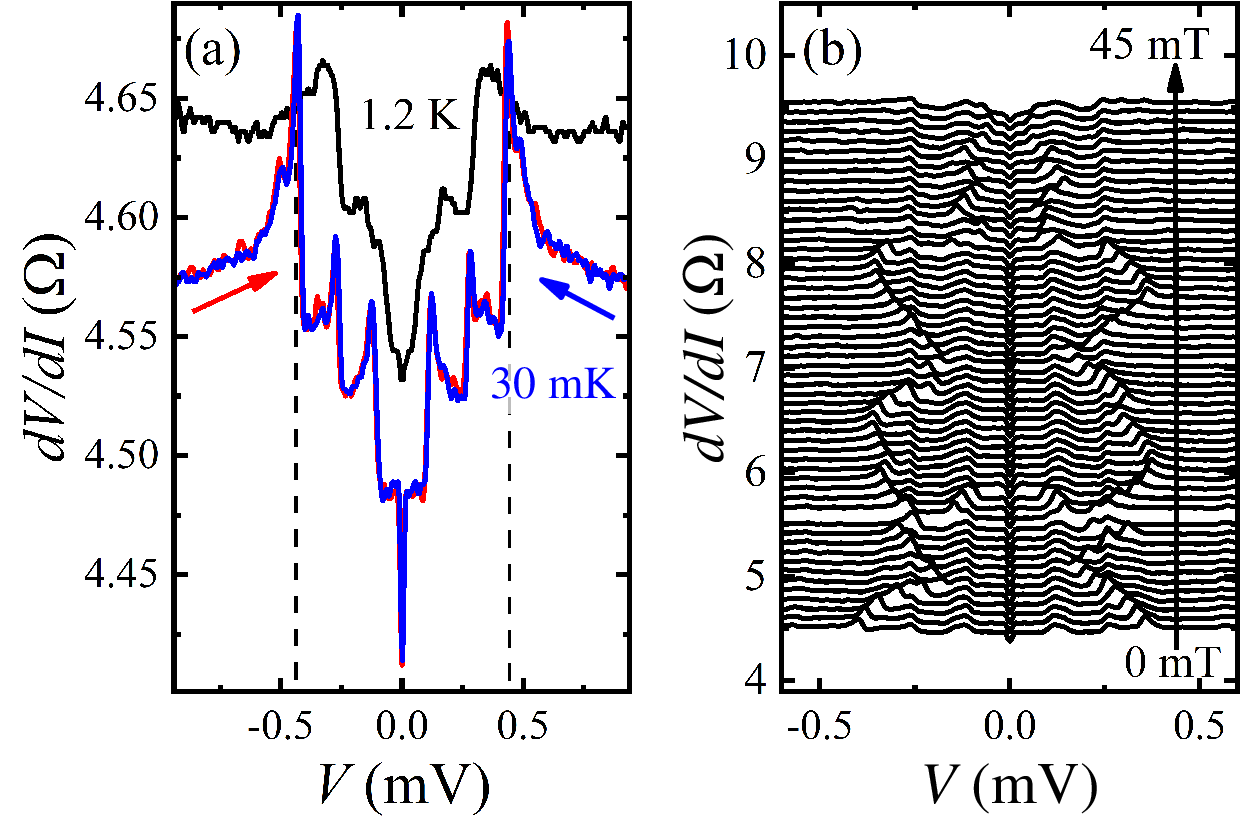}
\caption{(Color online) Andreev behavior of $dV/dI$ differential resistance for a single In-CrSb junction. 
(a) $dV/dI(V)$ curves in zero magnetic field at two different temperatures, 30~mK and 1.2~K, respectively. 
Differential resistance is diminished within $\approx\pm 0.5$~mV bias interval (depicted by vertical dashed 
lines). Temperature has low effect on  $dV/dI(V)$ curve, however, the zero-bias anomaly disappears at 1.2~K, as well 
as the fine subgap structures. 
(b) Magnetic field suppression of Andreev reflection as a waterfall plot (i.e., the $dV/dI(V)$ curves 
are shifted vertically). As expected, the superconducting gap is suppressed by magnetic field, but the 
suppression is non-monotonic: the gap oscillates with $\approx$13~mT period. The curves are obtained  at 30~mK temperature 
for the parallel to the In-CrSb interface magnetic field.
}
\label{fig4}
\end{figure}

As prepared, the In-CrSb interface transparency varies from junction to junction. For some neighbor In 
probes the superconducting order parameter is suppressed along the CrSb surface at distances smaller than $2 \mu$m probe separation. In this case, we can use the three-point connection scheme to investigate  differential resistance of a single (grounded)  In-CrSb junction, as described in the Samples section.

The typical $dV/dI(V)$ curve  is presented in Fig.~\ref{fig4} (a). We have verified that for a fixed grounded 
In contact,  the obtained $dV/dI(V)$ curves are independent of the mutual positions of current/voltage probes, so they 
indeed  reflect the resistance of  In-CrSb interface without any noticeable admixture of the 
CrSb bulk resistance. Due to these considerations, we should analyze Fig.~\ref{fig4} (a) in terms of Andreev 
reflection at single NS interface.

Since Andreev reflection allows subgap transport of Cooper pairs,  it appears experimentally as the resistance drop for voltages within the superconducting gap~\cite{andreev,tinkham}.  As it can be seen in Fig.~\ref{fig4} (a), differential resistance is reduced over a certain bias interval (see the vertical dashed lines), which is a bit smaller than the  known $0.5$~mV  bulk indium 
gap~\cite{indium-gap-field}. The partial gap suppression can be expected due to the finite spin polarization of the CrSb altermagnet surface, as it is supported by Josephson diode effect observation in Figs.~\ref{fig2} and~\ref{fig3}. 

The indium is always superconducting  for our 30~mK -- 1.2~K dilution fridge temperature range. Thus, temperature has low effect on the width of the differential resistance drop in Fig.~\ref{fig4} (a), the superconducting gap is only somewhat diminished at 1.2~K.  However,  the zero-bias anomaly disappears, as well as the fine subgap structures which are well developed at 30~mK.  

\begin{figure}
\includegraphics[width=\columnwidth]{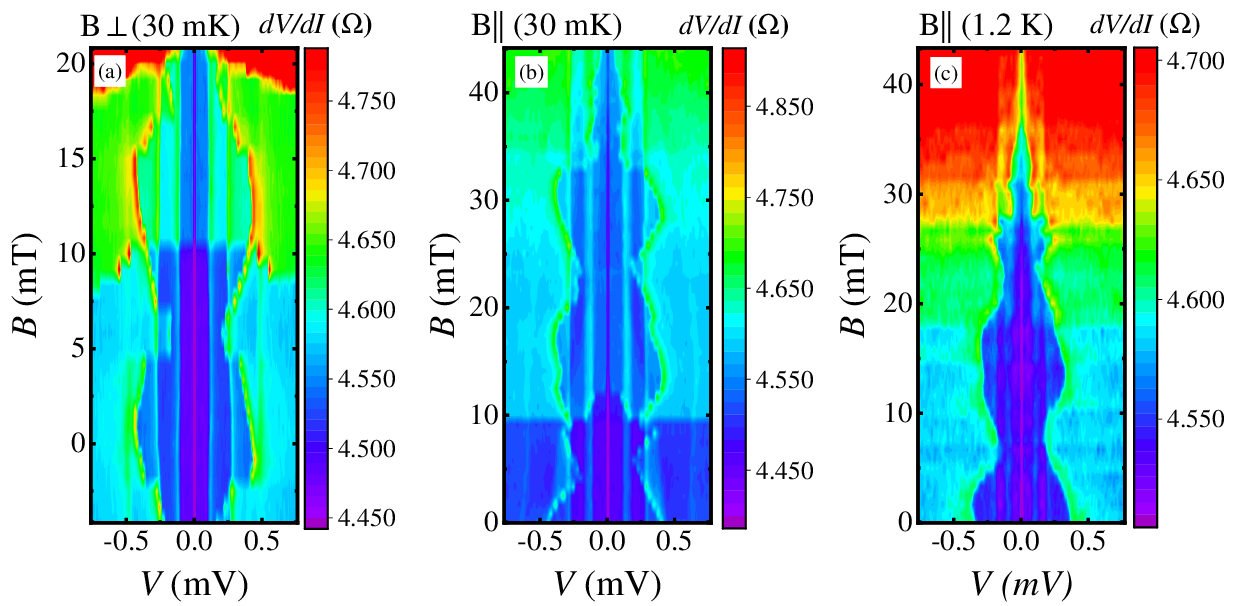}
\caption{(Color online) Non-monotonic behavior of the proximized superconductivity on the
surface of altermagnet CrSb as the detailed colormaps of $dV/dI(B,V)$ differential resistance at 30~mK temperature for 
two, normal (a) and in-plane (b), magnetic field orientations, respectively, and at 1.2~K for the in-plane 
field in (c). The superconducting gap oscillates for both field orientations with similar $\approx$13~mT 
period in (a) and (b), while it is suppressed earlier in normal to the In-CrSb interface magnetic field. The 
oscillations with the same period survive even at 1.2~K in (c).  
 }
\label{fig5}
\end{figure}

Since Andreev process is defined by the superconducting gap, it should be monotonically suppressed by magnetic field~\cite{andreev,tinkham}. The evolution of the  $dV/dI(V)$ curve in magnetic field is shown in  Fig.~\ref{fig4} (b) as a waterfall plot. The superconducting gap, determined as the full width of the differential resistance drop, is indeed suppressed by magnetic field, but the suppression is non-monotonic: the gap oscillates in the external field with $\approx$13~mT period, which well correlates with the shift of the zero-resistance region in Fig.~\ref{fig2} (b). In 
contrast, the subgap structures and the zero-bias anomaly, which are predicted for altermagnet-superconductor interface~\cite{subgap_AM}, are stable in magnetic field in Fig.~\ref{fig4} (b). 

The detailed picture of the gap suppression is shown in Fig.~\ref{fig5} (a-c) for two magnetic field 
orientations and two, 30~mK and 1.2~K, temperatures. The superconducting gap  oscillates for  both field 
orientations with similar $\approx$13~mT period (compare Figs.~\ref{fig5} (a) and (b)) while the gap is 
suppressed earlier in normal to the In-CrSb interface magnetic field. The oscillations with the same period survive even at 1.2~K in Fig.~\ref{fig5} (c) for the in-plane field orientation. It is natural to have the critical 
field anisotropy in Figs.~\ref{fig5} (a) and (b) for the  planar  geometry of our experiment, however, the suppression pattern demonstrates non-monotonic behavior of the proximized superconductivity on the surface of altermagnet CrSb.

\section{Discussion}

As a result of the experiment, we observe non-monotonic responses to an applied magnetic field in In-CrSb proximity devices, as well as a magnetic-field-induced asymmetry of the Josephson current in double In-CrSb-In junctions. In the latter case, the $dV/dI(B)$ curves are mirrored in respect to zero field for two magnetic field sweep directions, which is known for Josephson spin valves. For a single In-CrSb interface, the superconducting gap oscillates in magnetic field for both field orientations,  before its full suppression. 

Since indium is a conventional s-wave superconductor, the observed effects should mainly be associated with the specific properties of the proximized altermagnet CrSb. 

The $dV/dI(B)$ curves reversal in respect to the zero field is not expected for conventional Josephson junctions with a uniformly magnetized central layer, where remagnetization can, at most, shift the position of the $I_c(B)$ pattern in magnetic field~\cite{cro2,reverse}. We also verify, that there is no frozen flux in our solenoid for the field range in Fig.~\ref{fig2}. Moreover, the flux could only shift the zero-resistance region, therefore, it is inconsistent with the $dV/dI(B)$ curves reversal  in Fig.~\ref{fig2} and  the supeconducting gap oscillations in Figs.~\ref{fig4} and~\ref{fig5}.

By contrast, the observed behavior is a known fingerprint of Josephson spin valves~\cite{reverse,jsv1,jsv2,jsv3,jsv4,krasnov,jsv5,jsv6,jsv7}. 
Whereas in conventional Josephson junctions the supercurrent is primarily modulated by the magnetic flux through the junction area, in JSV it is largely defined by the relative orientation of magnetic layers, giving rise to the $I_c(B)$ asymmetry and reversal.

It seems to be important that altermagnet CrSb exhibits both altermagnetic 
and topological features, including  topological surface states~\cite{Weyl alter2_CrSb,Weyl alter1_CrSb,crsbsbs}.  Due to spin-momentum locking, the  topological surface states in CrSb are spin-polarized, in addition to the  
altermagnetic spin splitting in the bulk bands.  This naturally motivates a qualitative description of the In-CrSb-In junction in terms of the 
Josephson spin valve scenario with two distinct (surface and bulk) magnetic phases. In our experiment,  the strength of the  spin-valve effect depends on the $\pi/2$ rotation of the magnetic field in Figs.~\ref{fig2} and~\ref{fig3}, similarly to the interplay in magnetization  between the altermagnetic bulk and the topological surface contributions~\cite{crsbsbs}. Moreover, $dV/dI(B)$ hysteresis in Fig.~\ref{fig2} (a) is of approximately the same width as the unusual $M(H)$ diamagnetic hysteresis which has been attributed to the spin-polarized surface states~\cite{crsbsbs}.
  
The Josephson spin-valve behavior has recently been observed in Nb-Mn$_3$Ge-Nb junctions containing a single interlayer of the topological chiral antiferromagnet Mn$_3$Ge~\cite{jsv5}. The authors attributed this observation primarily to a proximity-induced, spin-polarised triplet supercurrent carried trough Mn$_3$Ge. It was argued that the  Berry-curvature-induced fictitious magnetic fields promote the spin-mixing and spin-rotation processes required for singlet-to-triplet pair conversion, thereby enabling long-range triplet supercurrent through a single chiral antiferromagnetic interlayer. At the same time, the role of topological surface states was not considered for short Nb-Mn$_3$Ge-Nb junctions~\cite{jsv5}. While we agree that proximity-induced spin-polarized triplet supercurrent is an  essential ingredient of the problem, we suggest that topological spin-polarized surface states  play a central role in establishing the Josephson spin-valve behavior for long (2~$\mu$m) In-CrSb-In junctions~\cite{topojj1,topojj2,topojj3,topojj4,topojj5}. 

The Josephson diode effect typically accompanies Josephson spin-valve behavior~\cite{infgt,JDE,aunite}, it can arise in 
superconductors hosting finite-momentum Cooper pairing~\cite{fmcp1,JDE,aunite}, which can be favorable in altermagnets under certain conditions~\cite{proxi2,AM_field_supercond,fmcp1,fmcp2,fmcp3}. 

The finite-momentum Cooper pairing can also be responsible for non-monotonic response to the applied magnetic field for a single In-CrSb junction in in Figs.~\ref{fig4},~\ref{fig5}. Among  various possible proximity-induced and intrinsic superconducting pairing states theoretically identified in altermagnets~\cite{alter_supercond_notes,alter_supercond_review1,proxi1,proxi2,proxi3,proxi4},
finite-momentum pairing may emerge and give rise to a reentrant superconducting state as a function of the
applied magnetic field~\cite{AM_field_supercond}. For example, a non-monotonic field dependence of the critical temperature occurs after 
the transition into the Fulde-Ferrell-Larkin-Ovchinnikov (FFLO) state~\cite{Mironovetal2021,Melnikovetal2022}. FFLO physics is based on finite-momentum Cooper pairing against a background of the Zeeman splitting, so it is fully compatible with the requirements for the Josephson diode effect~\cite{fmcp1,JDE,aunite}. This might be a reason to have similar $\approx$13~mT period in Figs.~\ref{fig4},~\ref{fig5} and $\approx$13~mT shift of the zero-resistance region in Fig.~\ref{fig2} (b).

\section{Conclusion}

As a conclusion, we experimentally investigated charge transport in In-CrSb and In-CrSb-In proximity devices, which are formed as junctions between superconducting indium leads and thick single crystal flakes of altermagnet CrSb. For double In-CrSb-In junctions, $dV/dI(B)$ curves are mirrored in respect to zero field for two magnetic field sweep directions, which is characteristic behavior of a Josephson spin valve. Also, we demonstrated Josephson diode effect by direct measurement of the critical current for two opposite directions in external magnetic field. We interpret these observations as a joint effect of the spin-polarized topological surface states and the altermagnetic spin splitting of the bulk bands in CrSb. For a single In-CrSb interface, the superconducting gap oscillates in magnetic field for both field orientations, which strongly resembles the Fulde-Ferrell-Larkin-Ovchinnikov behavior. FFLO is based on finite-momentum Cooper pairing, therefore, it is fully compatible with the requirements for the Josephson diode effect.

\acknowledgments
We wish to thank S.S.~Khasanov for X-ray sample characterization.

\end{document}